\documentstyle[stwol,epsfig]{article}

\def\Journal#1#2#3#4{{#1} {\bf #2}, #3 (#4)}

\def\NPB{{\em Nucl. Phys.} B}
\def\PLB{{\em Phys. Lett.}  B}

\def\ZPC{{\em Z. Phys.} C}
\def\CPC{\em Comp. Phys. Comm.}

\def\mco{\multicolumn}
\def\mcotc{\multicolumn{2}{|c|}}

\newcommand\mr[1]{{\mbox{\rm #1}} }
\newcommand{\epem}{e$^+$e$^-$}

\newcommand\II{\ifmmode \ell\ell \else $\ell\ell$\fi}
\newcommand\lplp{\ifmmode \ell^\prime\ell^\prime \else $\ell^\prime\ell^\prime$\fi}
\newcommand\qq{\ifmmode \mr q \mr q \else qq\fi}
\newcommand\ff{\ifmmode \mr f \mr f \else ff\fi}
\newcommand\ee{\ifmmode \mr e \mr e \else ee\fi}
\newcommand\mm{\ifmmode \mu\mu \else $\mu\mu$\fi}
\newcommand\tata{\ifmmode \tau\tau \else $\tau\tau$\fi}
\newcommand\nn{\ifmmode \nu\nu \else $\nu\nu$\fi}
\newcommand\nnV{\ifmmode \nu\nu\mr V \else $\nu\nu$V\fi}
\newcommand\llV{\ifmmode \ell\ell\mr V \else $\ell\ell$V\fi}
\newcommand\ttV{\ifmmode \tau\tau\mr V \else $\tau\tau$V\fi}
\newcommand\lnqq{\ifmmode \ell\nu\mr q \mr q \else $\ell\nu$qq\fi}
\newcommand\llll{\II\lplp}
\newcommand\llqq{\II\qq}
\newcommand\nnll{\nn\II}
\newcommand\nnee{\nn\ee}
\newcommand\nnff{\nn\ff}
\newcommand\nnmm{\nn\mm}
\newcommand\nntt{\nn\tata}
\newcommand\nnqq{\nn\qq}
\newcommand\eeqq{\ee\qq}
\newcommand\eeff{\ee\ff}
\newcommand\mmqq{\mm\qq}
\newcommand\ttqq{\tata\qq}
\newcommand\qqbar{\ifmmode \mr q\overline{\mr q} \else q$\overline{\mr q}$\fi}
\newcommand\ffbar{\ifmmode \mr f\overline{\mr f} \else f$\overline{\mr f}$\fi}
\newcommand\eetoqq{\ifmmode \epeme\to\qqbar \else \epem$\to$\qqbar\fi}
\newcommand\sr[1]{{\mbox{\scriptsize\rm #1}} }
\newcommand\mZ{\ifmmode m_{\sr Z} \else $m_{\sr Z}$\fi}
\newcommand\sqrts{\ifmmode \sqrt{s} \else $\sqrt{s}$\fi}

\def\al{\alpha}

\def\beq{\begin{equation}}
\def\eeq{\end{equation}}
\def\bea{\begin{eqnarray}}
\def\eea{\end{eqnarray}}

\begin{document}




\renewcommand{\thefootnote}{\fnsymbol{footnote}}

\title{MEASUREMENTS OF FOUR-FERMION PRODUCTION\\ VIA NEUTRAL ELECTROWEAK CURRENTS AT LEP \footnotemark[1]}
\thispagestyle{myheadings}
\markright{FREIBURG-EHEP-96-02, OPAL CR-297}

\author{\vskip -2.5 mm Michael KOBEL}

\address{Fakult\"at f\"ur Physik, Universit\"at Freiburg,
Hermann-Herder-Str.~3, D-79104 Freiburg, FRGermany\\
\rm representing the four LEP collaborations}


\twocolumn[\maketitle\abstracts{
Four-fermion production via electroweak neutral currents has been
measured by all four LEP collaborations at center-of-mass energies
near the Z resonance and for the first time also at energies
well above the Z peak. Essentially all possible final states
have been covered in four different topologies. 
}]

\section{Introduction}

\vspace*{-0.1cm}
The study of four-fermion production via neutral electroweak currents
at LEP is one of the basic
tests of higher order processes in the electroweak Standard Model.
Theoretically 
this process is well understood within the Standard Model, 
and experimentally four-fermion events have a clear signature.
Neutral current four-fermion production also constitutes  an important background for
particle searches, so that
specialized analyses, like searches for Higgs or SUSY particles, try to suppress
it. A dedicated study of neutral current
four-fermion production thus  represents a complementary search
for new physics, that could manifest itself
in sizable deviations of rates or kinematic distributions
from the expectations.
\footnote{
Invited talk given at the 28th International Conference on High Energy Physics,
ICHEP96, 25-31 July, 1996, Warsaw, Poland; to be published in the proceedings.}

\begin{figure}[bht]
\begin{picture}(50,70)(0,7)
\epsfysize=2.8cm
\epsffile{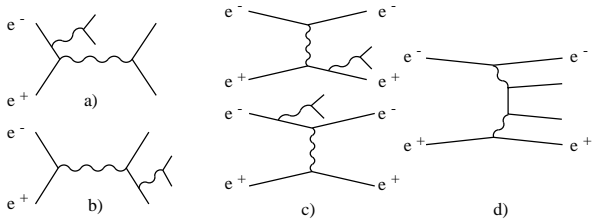}
\end{picture}
\vspace{0cm}\caption[]{Feynman diagrams for four-fermion final states
involving only neutral gauge boson exchange, represented by wiggly lines.
The solid lines not labelled as e$^\pm$ may be quarks, charged leptons,
or neutrinos in case of Z exchange: a) ``conversion'', b) ``annihilation'',
c) ``bremsstrahlung'', and d) multiperipheral production. 
}
\label{diagrams}
\vspace*{-0.35cm}
\end{figure}

In the following we review results from all LEP collaborations
obtained for the final states \llll, \llqq, \nnll, and \nnqq,
($\ell$ = e, $\mu$, $\tau$) at center-of-mass energies, \sqrts,
near the Z resonance (LEP-1), at \sqrts\ = 130--136~GeV (LEP~1.5), and
from the ongoing LEP-2 run at \sqrts\ = 161~GeV.
The above channels cover all experimentally accessible four-fermion
final states that proceed via neutral gauge boson exchange according to the
diagrams in Fig.~\ref{diagrams}. The only exception is the 
qqqq final state that is difficult to distinguish from higher order
QCD corrections to \eetoqq.
The relative contributions from the various diagrams vary strongly
with \sqrts, and are especially large if one of the intermediate bosons
is close to its mass shell.
At the highest center-of-mass energies, above the
W$^+$W$^-$ threshold, there are  irreducible contributions
from charged current diagrams, notably to the \nnll\ final state.

\vspace*{-0.2cm}
\section{Event selection}
\vspace*{-0.1cm}

In the following we describe the main features
of the event selections that are largely in common for
all four experiments. In detail, the  signal acceptances
and background contaminations can still be rather different
amongst the experiments, as will be obvious from the results
 in Section~\ref{results}.

Cuts on two quantities separate four different event classes, which
will be dicussed in the following subsections.
First, cuts on missing momentum separate channels with and without neutrinos.
These are then further subdivided by cuts the charged particle multiplicity.
The \llll\ and \nnll\ channels feed the low multiplicity classes, 
while the \llqq\ and \nnqq\ channels contribute to both high and low
multiplicity final states, depending on the invariant qq mass available
for fragmentation.

\vspace*{-0.2cm}
\subsection{Low multiplicity and large missing momentum: \nnV}

With a multiplicity requirement of two,
this class contains mainly \nnee\ and \nnmm\
events. To optimize the
efficiency, not all experiments perform a detailed
lepton identification.
These selections then also accept \nntt\ events as well as
\nnqq\ channels where the quarks fragment to two charged hadrons,
mainly via intermediate vector resonances like $\rho,\omega$, and $\phi$.
The (usually low mass) lepton or hadron pair is denoted as V in the following.
Further kinematical cuts are applied to reduce the
main background sources, namely  lepton pairs from two-photon processes,
and $\gamma$\nn\ events with a converted photon.

\vspace*{-0.2cm}
\subsection{High multiplicity and large missing momentum: \nnqq}

This class is composed of  \nnqq\ states with a charged 
multiplicity of 4 or more. Again kinematical cuts,
like on the missing transverse momentum, reject the two-photon
background, whereas \eetoqq\ background is reduced by requiring a
large missing mass. Events where the missing momentum
is carried by the decay products of a heavy particle, like the Z$^0$
in the Standard Model ``conversion'' process (Fig.~\ref{diagrams}a),
or by (new) invisible heavy particles,
will therefore pass this selection.

\vspace*{-0.2cm}
\subsection{Low multiplicity and small missing momentum: \llV}

This channel covers the multiplicities four and six, to allow
for \ttV\ contributions with one 3-prong $\tau$ decay, where again V
stands for a lepton or hadron pair.
While for the LEP-1 data, the various two-pair combinations
contributing to this class are further subdivided 
into electron, muon, tau, and hadron pairs
by means of lepton identification,
this is usually not the case for the low-statistics
measurements at  \sqrts\ well above the Z resonance.
A rejection of photon conversions is performed to reduce
the background from $\gamma$ff events to the \eeff\ channel,
and 3-prong mass cuts are applied to suppress the \tata\ background.

\vspace*{-0.2cm}
\subsection{High multiplicity and small missing momentum: \llqq}

\begin{figure}[htb]
\begin{picture}(60,170)(6,20)
\epsfysize=8.5cm
\epsffile{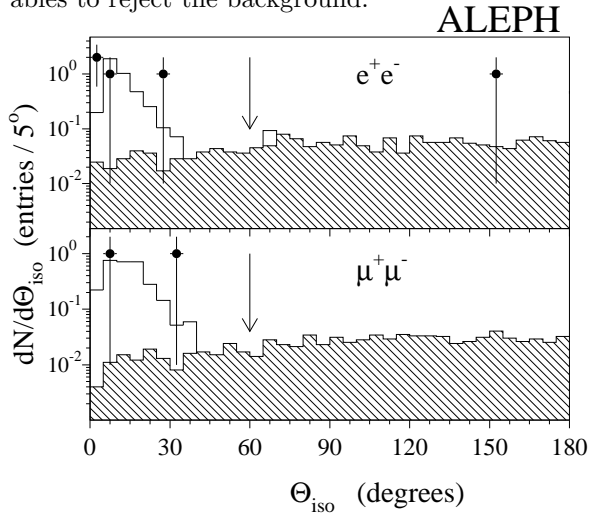}
\end{picture}
\vspace{0cm}\caption[]{The sum, $\Theta_{\rm iso}$, of the individual lepton
isolation angles as measured by ALEPH
for \eeqq\ and \mmqq\ candidate events at \sqrts\ = 130--136~GeV. All cuts,
except isolation, have been applied. The hatched histograms show the signal
expectation, the solid one the background, data are the points, and the arrow indicates
the cut value.
}
\label{alephiso}
\vspace*{-1.0cm}
\end{figure}

In this channel  \eetoqq\ and, above the W$^+$W$^-$
threshold, also the \lnqq\ semileptonic W-pair decays are the main backgrounds.
Therefore an explicit lepton identification is necessary in order 
to avoid, that split-off tracks from quark jets are taken as lepton candidates.
Results are therefore given separately for the \eeqq\ and \mmqq\ channel.
The Standard Model \ttqq\ expectation is very small.
Minimum lepton momentum requirements and especially  the isolation of the lepton pair 
from the quark jets (Fig.~\ref{alephiso}) are the most powerful variables
to reject the background.

\section{Results}
\label{results}
\vspace*{-0.2cm}

\begin{table*}[tb]
\caption{LEP-1 results for four-fermion analyses. The upper line gives the numbers of events
expected for Standard Model signal + background, the lower line the experimental observations.
See the text for typical errors of the signal and background expectations.
A background of 0 means, that the expectation is below 0.5 events. 
The entry \nnff\ is the sum over \nnV\ (1 event observed) and \nnqq\ (2 events observed).
\label{lep1res}}
\vspace{0.4cm}
\begin{center}
\begin{tabular}{|l|c||l@{ + }r|l@{ + }r|l@{ + }r|l@{ + }r||l@{ + }r|}
\hline
\sqrts$\approx$\mZ  &{$\int\cal L$}dt
                    &\mcotc{\nnff}&\mcotc{\llV}&\mcotc{\eeqq}&\mco{2}{|c||}{\mmqq}&\mcotc{sum} \\
\hline\hline
ALEPH~\cite{Alep1}  
          & 79~pb$^{-1}$& 3   &  0    & 232 & $<2$ &  14 &     1 &  13 &     0        & 262 &$<3$  \\ 
                 &      &  \mcotc{3}  & \mcotc{229}& \mcotc{10}  &\mco{2}{|c||}{19}   & \mcotc{261}\\
\hline
DELPHI~\cite{Dlep1} 
          & 16~pb$^{-1}$& --   &  --  & 23 &  3    &  -- &    -- &  -- &     --       &  23 &  3   \\ 
                 &      &  \mcotc{--} & \mcotc{27} & \mcotc{--}  &\mco{2}{|c||}{--}   & \mcotc{27 }\\
\hline
L3~\cite{Llep1} 
          & 36~pb$^{-1}$& --   &  --  & 45 &  4    &  15 &     1 &   5 &     0        & 65  &  5   \\ 
                 &      &  \mcotc{--} & \mcotc{43} & \mcotc{18}  &\mco{2}{|c||}{6}    & \mcotc{67} \\
\hline
OPAL~\cite{Olep1,Olep15} 
          &132~pb$^{-1}$& --   &  --  & 44 & 13    &  19 &     1 &  20 &     8        & 83  & 22   \\ 
 & (\llV: 19~pb$^{-1}$) &  \mcotc{--} & \mcotc{50} & \mcotc{25}  &\mco{2}{|c||}{28}   & \mcotc{103}\\
\hline
\end{tabular}
\end{center}
\end{table*}

The selected number of events in the data,  
and their kinematical distributions
have been compared with the expectations for signal and background events. 
In the following we will tabularize the number of expected and observed
events for the different classes at the various values of \sqrts.
For better readability of the tables, the errors on these numbers
have been omitted. 
All results for LEP-1 are final, while for LEP-1.5
only the OPAL collaboration (for \eeqq\ and \mmqq) and the ALEPH collaboration
have published their results. All other results at LEP-1.5 and LEP-2 are preliminary.

The signal predictions are mostly derived
from the {\tt FERMISV}~\cite{fsv} generator, usually after corrections
for dominant higher order effects (e.g. running $\al$ and vector resonances, typically amounting
to 10--15\%) have been applied. Residual experimental and theoretical 
uncertainties of the signal predictions are of the order of 10\%.
Expected backgrounds are usually small but, especially above the Z resonance, 
the precision of the background prediction suffers from low statistics
and the modelling of rare phase space regions 
and can in some cases reach uncertainties of a factor of two.

\vspace*{-0.2cm}
\subsection*{3.1~~Results at $\sqrts\approx\mZ$\ (LEP-1)}

At \sqrts\ near the Z resonance the dominant four-fermion process
is the final state radiation of a fermion pair from one of the
decay products of the intermediate Z (Fig.~1b), i.e. from a lepton in \llV\
events and (preferrably) a quark in \llqq\ events. 
Initial state fermion pair radiation for a {\it on-shell} intermediate Z
(Fig 1a) is suppressed by the available phase space. 
This leads to a suppression of the \nnff\ topology, since here Fig.~\ref{diagrams}a
is the only neutral current diagram contributing.

The four LEP experiments, based on different amounts of integrated luminosity
{$\int\cal L$}dt, have covered a large variety of channels, as detailed
in Table~\ref{lep1res}. 
No deviations from expected cross-sections have been found
with a statistics of typically 3 \nnff, 20-50 \llqq, and 30-230 \llV\ events for each 
experiment. An early excess of \tata V events observed by ALEPH~\cite{Alep0}
was not confirmed with larger statistics. Likewise all kinematical distributions
agree with the expectations. One exception is a slightly unlikely 
mass and transverse momentum configuration of the three ALEPH \nnff\ events,
which becomes less significant 
after taking into account also contributions from charged current exchanges~\cite{Alepc}.

\vspace*{-0.2cm}
\subsection*{3.2~~Results at \sqrts\ = 130--136~GeV (LEP-1.5)}
At center-of-mass energies well above the Z resonance the dominant
four-fermion process, apart from the 
$t$-channel ``bremsstrahlung'' diagram for  the \eeff\ final state,
is the initial state emission of a
fermion pair with an energy appropriate for  
a ``return'' to the Z resonance. 
This leads to a clear separation of the two fermion pairs and increases
the selection efficiency, especially for the \llqq\ channels,
resulting in a visible cross-section comparable to that on the
Z resonance.

\begin{table*}[tb]
\caption[]{LEP-1.5 results for four-fermion analyses. The upper line gives the numbers of events
expected for Standard Model signal + background, the lower line the experimental observations.
See the text for typical errors of the signal and background expectations.
A background of 0.0 means, that the expectation is below 0.05 events.
\vspace*{-0.2cm}
\label{lep15res}}
\vspace{0.4cm}
\begin{center}
\begin{tabular}{|l|c||l@{ + }r|l@{ + }r|l@{ + }r|l@{ + }r|l@{ + }r||l@{ + }r|}
\hline
\sqrts$\approx$133~GeV &{$\int\cal L$}dt
                    &\mcotc{\nnV}&\mcotc{\nnqq}&\mcotc{\llV}&\mcotc{\eeqq}&\mco{2}{|c||}{\mmqq}&\mcotc{sum} \\
\hline\hline
ALEPH~\cite{Alep15} 
& 5.8~pb$^{-1}$& 1.8 & 0.0 &  0.5 & 0.0  &  1.9 & 0.0 &  1.7 &  0.1 &  0.9 & 0.0         & 6.7 & 0.2  \\ 
        &      & \mcotc{2} &  \mcotc{1}  &  \mcotc{1} &  \mcotc{1}  & \mco{2}{|c||}{0}   & \mcotc{5}  \\
\hline
DELPHI~\cite{Dlep15}
& 5.9~pb$^{-1}$& 0.3 & 0.0 &  0.3 & 0.3  &  0.2 & 0.0 &  0.9 &  0.3 &  0.7 & 0.1         & 2.4 & 0.7  \\ 
        &      & \mcotc{1} &  \mcotc{0}  &  \mcotc{0} &  \mcotc{0}  & \mco{2}{|c||}{2}   & \mcotc{3}  \\
\hline
L3~\cite{Llep15}
& 5.0~pb$^{-1}$& --  & --  &  --  & --   &  --  & --  &  1.4 &  0.7 &  0.5 & 0.0         & 1.9 & 0.7  \\ 
        &      & \mcotc{--}&  \mcotc{--} &  \mcotc{--}&  \mcotc{2}  & \mco{2}{|c||}{0}   & \mcotc{2}  \\
\hline
OPAL~\cite{Olep15,OlepV} 
& 5.2~pb$^{-1}$ & --  & --  &  --  & --   &  1.3 & 0.1 &  0.6 &  0.1 &  0.5 & 0.1         & 2.5 & 0.3  \\ 
         &      & \mcotc{--}&  \mcotc{--} &  \mcotc{2} &  \mcotc{1}  & \mco{2}{|c||}{5}   & \mcotc{8}  \\
\hline\hline
LEP-1.5 & 21.9~pb$^{-1}$& 2.0 & 0.1&  0.8 & 0.3  &  3.4 & 0.2 &  4.7 &  1.2 &  2.6 & 0.2         & 13.5 & 1.9  \\ 
                &      & \mcotc{3} &  \mcotc{1}  &  \mcotc{3} &  \mcotc{4}  & \mco{2}{|c||}{7}   & \mcotc{18}  \\
\hline
\end{tabular}
\end{center}
\end{table*}

\begin{figure}[h]
\begin{picture}(60,140)(6,10)
\epsfysize=5.4cm
\epsffile{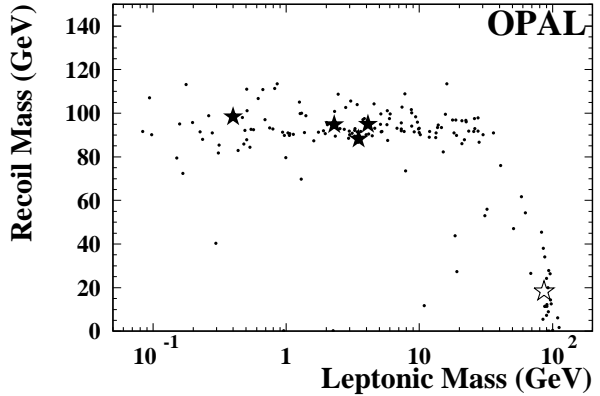}
\end{picture}
\vspace{0cm}\caption[]{Scatter plot of the mass of the lepton pair
and its recoil mass for the \mmqq\ channel at 130-136 GeV,
as observed by OPAL.
The small points is the {\tt FERMISV} expectation for the signal
distribution. The four data events with large recoil mass are
indicated by filled stars, whereas the open star represents a
low recoil mass event.
}
\label{opal_llqq}
\vspace*{-0.5cm}
\end{figure}

The results of the four LEP experiments at \sqrts=130--136~GeV are summarized
in Table~\ref{lep15res}. Generally, the observed numbers of events agree
well with the expectations. The only exception is the \mmqq\ channel,
where the OPAL collaboration observes 5 events for 0.55 signal and 0.06
background events expected~\cite{Olep15}.
This excess is not seen by the three other collaborations, that together observe 2 events
in this channel, where 2.2 are expected. Apart from their abundance, all
other features of the OPAL \mmqq\ events are consistent with the Standard Model expectation,
as is illustrated for the invariant mass distributions in Fig.~\ref{opal_llqq}.

Summing all final states from all collaborations at LEP-1.5 we observe 
18 neutral current four-fermion candidate events
in a total of 22~pb$^{-1}$ data luminosity.
Since 13.5 signal and 1.9 background events are predicted,
there is no indication of an additional strong source for
these final states within or beyond the Standard Model.
 
\vspace*{-0.2cm}
\subsection*{3.3~~Results at \sqrts\ = 161~GeV (LEP-2)}

The analysis of the same channels as at LEP-1.5 has been repeated by the
four LEP collaborations with similar cuts also for the ongoing run at
\sqrts\ = 161~GeV, that has started two weeks before the start of this conference.
In the first 10 days of running a sum of about
11~pb$^{-1}$ integrated luminosity has been collected by the four
experiments together. Three four-fermion events have been observed
(1 \eeqq\ by OPAL, 1 \mmqq\ by DELPHI, and 1 \nnqq\ by DELPHI),
compared to an expectation of 3.9 signal and 0.8 background events~\cite{Dlep15,Olep2,Alep2,Llep2},
summed over all channels and experiments apart from the three DELPHI channels
without observed events.

\vspace*{-0.3cm}
\section{Conclusions}
\vspace*{-0.1cm}

Four-fermion production via electroweak neutral currents has been
measured by all four LEP collaborations at center-of-mass energies
near the Z resonance and for the first time also at energies
well above the Z peak where different diagrams dominate.
Deviations from the
Standard Model predictions in single channels by single experiments
are consistent with statistical fluctuations.
At none of the center-of-mass energies
there is  evidence for a yet unknown extra source of four-fermion events.

\vspace*{-0.3cm}
\section*{Acknowledgments}
\vspace*{-0.1cm}
I would like to thank S.~Gonzales and Ch.~Hoffmann (ALEPH),
A.~Lipniacka and J.~Marco (DELPHI), and F.~Di~Lodovico (L3) for
supplying me with the most recent results and plots from the ongoing 161~GeV
run, and for helpful discussions.


\vspace*{-0.3cm}
\section*{References}
\begin{thebibliography}{99}
{\small 
\bibitem{fsv}  J.~Hilgart, R.~Kleiss and F. Le Diberder,\\ 
\Journal{\CPC}{75}{191}{1993}.

\bibitem{Alep1} ALEPH Collab., 
\Journal{\ZPC}{66}{3}{1995}.

\bibitem{Dlep1} DELPHI Collab., 
\Journal{\NPB}{403}{3}{1993}.

\bibitem{Llep1} L3 Collab., 
\Journal{\PLB}{321}{283}{1994}.

\bibitem{Olep1} OPAL Collab., 
\Journal{\PLB}{287}{389}{1992}.

\bibitem{Olep15} OPAL Collab., 
\Journal{\PLB}{376}{315}{1996}.

\bibitem{Alep0} ALEPH Collab., 
\Journal{\PLB}{263}{112}{1991}.

\bibitem{Alepc} ALEPH Collab., 
\Journal{\PLB}{334}{244}{1994}.

\bibitem{Alep15} ALEPH Collab., 
CERN-PPE/96-088, June 1996; submitted to \PLB.

\bibitem{Dlep15} DELPHI Internal Note ``Search for four-fermion final states
        in \epem\ interactions at \sqrts\ = 130--136~GeV'', update of 23 July 1996.

\bibitem{Llep15} L3 Internal Note 1959, May 1996.

\bibitem{OlepV} OPAL Physics Note 228, 19 July 1996;\\ 
contributed paper to ICHEP Warsaw, PA07-27. 

\bibitem{Olep2} OPAL Physics Note 239, 23 July 1996;\\
contributed paper to ICHEP Warsaw, PA07-25. 

\bibitem{Alep2} Christian Hoffmann (ALEPH),\\ private communication, 23 July 1996. 

\bibitem{Llep2} Francesca~Di~Lodovico (L3),\\ private communication. 23 July 1996. 
}
\end{thebibliography}
\end{document}